\documentclass[%
 reprint,
 amsmath,amssymb,
 aps,
]{revtex4-2}

\usepackage{graphicx}% Include figure files
\usepackage{dcolumn}% Align table columns on decimal point
\usepackage{bm}% bold math
\usepackage{url}
 
\usepackage{babel}   % multi-language support
\usepackage{float}   % floats
\usepackage{url}     % urls
\usepackage{graphicx}% Include figure files
\usepackage{layout}
\usepackage{amsmath}
\usepackage{tabularx}
\usepackage{cleveref}

\usepackage[left=1cm,right=2.5cm,top=2cm,bottom=1.5cm]{geometry}

\begin{document}

\preprint{APS/123-QED}

\title{Universal polar instability in highly orthorhombic perovskites }

\author{Cameron A.M. Scott}
\email{cameron.a.scott@durham.ac.uk}
\author{Nicholas C. Bristowe}%
 \email{nicholas.bristowe@durham.ac.uk}
\affiliation{%
 Centre for Materials Physics, Durham University, South Road, Durham DH1 3LE, United Kingdom\\
}%

\begin{abstract}
The design of novel multiferroic ABO$_3$ perovskites is complicated by the presence of necessary magnetic cations and ubiquitous antiferrodistortive modes, both of which suppress polar distortions. Using first-principles simulations, we observe that the existence of quadlinear and trilinear invariants in the free energy, coupling tilts and antipolar motions of the A and B sites to the polar mode, drives an avalanche-like transition to a non-centrosymmetric $Pna2_1$ symmetry in a wide range of magnetic perovskites with small tolerance factors - overcoming the above restrictions. We find that the $Pna2_1$ phase is especially favoured with tensile epitaxial strain, leading to an unexpected but technologically useful out-of-plane polarization. We use this mechanism to predict various novel multiferroics displaying interesting magnetoelectric properties with small polarization switching barriers.

\end{abstract}

\maketitle

Perovskite materials with the ABX$_3$ chemical formula provide a fascinating playground for exploring the physics of transition metal compounds. Diverse phenomena including non-collinear magnetism, metal-insulator transitions, (anti-)ferroelectricity and superconductivity are known to exists in the structure type. Such a wide range of physical phenomena is enabled by the ability of the structure to distort - largely through tilts and rotations of the B-site octahedra - to accomodate almost any element on the A and B sites. For a given ABX$_3$ composition, the tendency towards distortion is phenomenologically described by the tolerance factor,
\begin{equation}
t = \frac{r_\textrm{A}+r_\textrm{X}}{\sqrt{2}(r_\textrm{B}+r_\textrm{X})}
\end{equation}
where $r_\textrm{A}$, $r_\textrm{B}$ and $r_\textrm{X}$ are the ionic radii of the A, B and X sites respectively. For small variations from the ideal $t=1$, octahedral distortions occur to alleviate the size mismatch. For very small or very large $t$ (typically $t<0.7$ or $t>1.0$), distortions can not stabilize the perovskite structure and other structural polymorphs are favoured.
Such structural and chemical flexibility make perovskites ideal for engineering desired physical properties. However, engineering multiferroism (the simultaneous ordering of both electric and magnetic dipoles which can be reversed by applied external fields) with a strong coupling between the constituent electric and magnetic dipole orders has proved challenging. There are two main reasons for this: 1) polar distortions in perovskites are typically caused by the formation of bonds between B and X sites \cite{cohen1992origin}. If the B-site possesses the localized $d$ electrons necessary for long-range magnetic ordering, such bonds are suppressed by the $d^0$ criterion \cite{hill2000there} and 2) the octahedral tilts necessary to accommodate a wide range of cations suppress polar distortions \cite{zhong1995competing,benedek2013there}. 

Despite the limitations enforced by the above phenomena, various mechanisms have been identified that result in the coexistence of ferromagnetism and ferroelectricity in a perovskite architecture. For example, lone-pair cations on the A-site separate the magnetic and polar modes onto distinct cations  \cite{kimura2003magnetocapacitance,teague1970dielectric} whilst layered perovskites can break centrosymmetry and allow ferroelectricity in a hybrid improper mechanism  \cite{bousquet2008improper,benedek2011hybrid}. Alternatively, the strong coupling of epitaxial strain to the polar $\Gamma$-point phonon \cite{schlom2007strain} can result in strain-stabilized non-centrosymmetric phases. This distortion is even seen to occur in systems possessing $d$ electrons and thus provides a mechanism to circumvent the tendency of $d^n$ systems to form centrosymmetric structures \cite{bousquet2011induced,yang2012revisiting}. 

In this paper, we use first principles simulations to explore an alternative mechanism to engineer multiferroicity in a wide range of perovskites despite the presence of magnetic cations and large octahedral tilts. Specifically, we observe that couplings of the polar distortion and antipolar motions of B-site cations to various antiferrodistortive modes such as octahedral rotations stabilizes an out-of-plane polarization through an unusual avalanche-related mechanism \cite{perez2008multiple}. Furthermore, this mechanism appears to be universal to all perovskites with large tilting and does not rely on lone-pairs or $d^0$ ions (or layering to break centrosymmetry). While not fundamentally reliant on tensile strain, the mechanism is substantially enhanced by its application due to how strain influences the antiferrodistortive modes present in the couplings. Unexpectedly, tensile strain leads to strong out-of-plane polarizations in sharp distinction to previous studies of strain-induced ferroelectricity \cite{rabe2005theoretical}. Finally, we show how these principles can be used to create promising multiferroics with strong spin-phonon couplings and illustrate these effects on a few candidate materials.

\section{RESULTS}
\subsection{\label{generality}Origins Of \textit{P\lowercase{na}$2_1$} Instability}
Previous work \cite{belik2014high,belik2012crystal,ding2017unusual,bosak2001epitaxial,salvador1998stabilization,li2012polar} has demonstrated that perovskites with a low tolerance factor can be stabilised under high-pressure or thin film synthesis techniques and very often form GdFeO$_3$ type structures with a $Pnma$ symmetry. This phase consists of two primary distortions from the cubic $Pm\bar{3}m$ reference -  equal antiphase tilts about two of the axes of the octahedra and an in phase tilt of a different magnitude about the final axis, a tilt pattern described symbolically in Glazer notation as $a^-b^+a^-$  \cite{glazer1972classification}. This results in a cell that is $\sqrt{2}{\times}2{\times}\sqrt{2}$ larger than $Pm\bar{3}m$. 

We explore instabilities of the $Pnma$ phase in a wide range of perovskite materials (see Table \ref{tab:parameters} and caption for details) as a function of epitaxial strain using first-principles simulations (see Methods). We choose to disentangle the effects of lone-pair stereochemistry and magnetically driven symmetry lowering by choosing materials in which the A-site does not possess lone pairs and is also non-magnetic ($f$ electrons were frozen to the core for rare earth cations). To demonstrate the generality of the effect, we also chose to study a selection of chemical compositions covering elements from the $s-$, $p-$, $d-$ and $f-$ blocks with a range of formal valences and $d$ occupancies. 

\begin{figure*}[t]
\centering
\includegraphics[width=\textwidth]{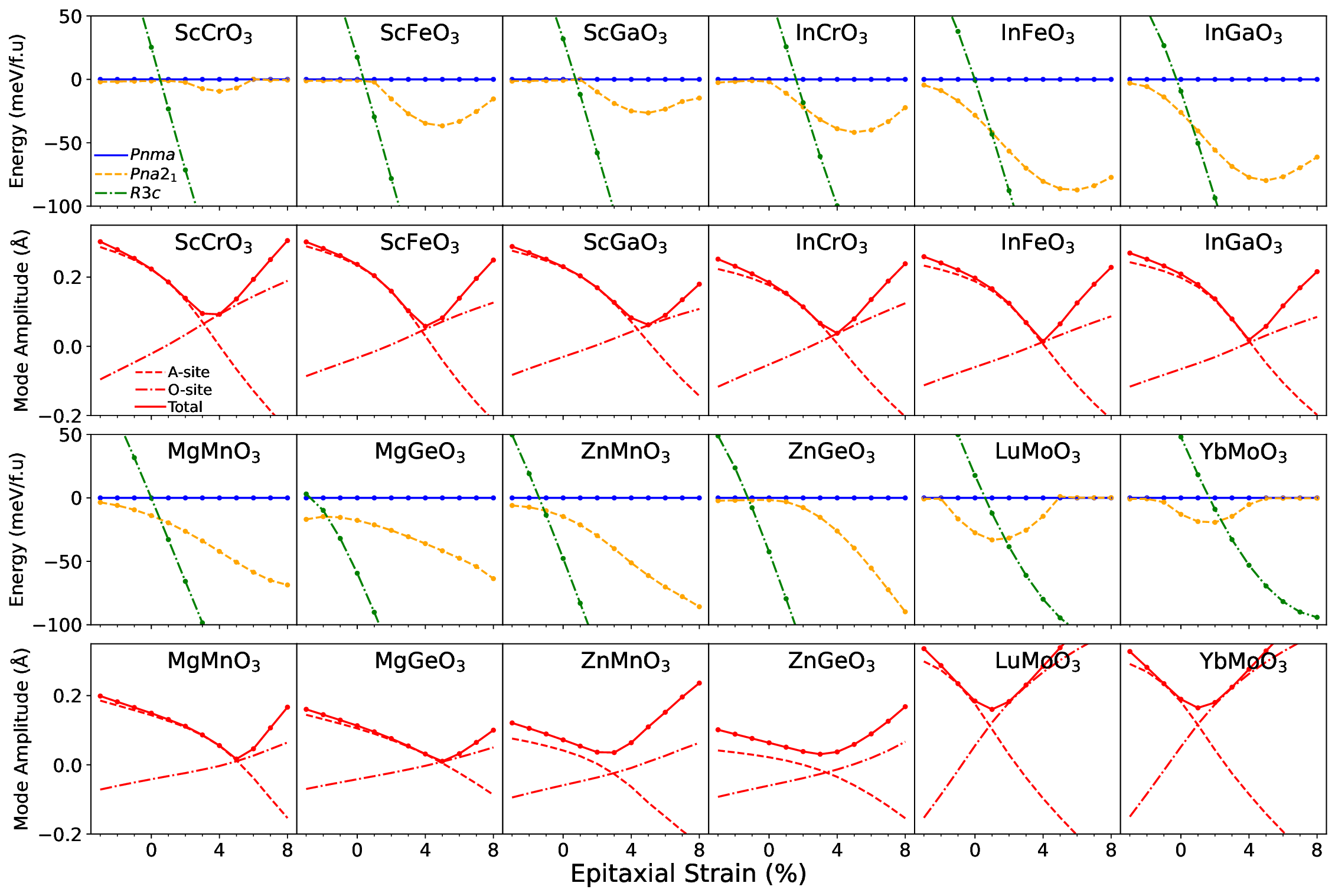}% Here is how to import EPS art
\caption{\label{fig:energy-strain} First and third rows : energy of the $Pna2_1$ (orange dashed) and $R3c$ (green dash-dot) phases, relative to $Pnma$ (blue solid), vs epitaxial strain for a variety of oxide perovskites with small tolerance factors.  Second and fourth rows : Total $R_4^-$ amplitude as well as the $R_4^-$ amplitude decomposed into the components affecting the A and O sites individually. Details of simulations and lattice constants can be found in Table \ref{tab:parameters}. }
\end{figure*}

Despite the large antiferrodistortive modes present in these highly orthorhombic materials, which are known to strongly suppress any polar distortions \cite{benedek2013there}, the first and third rows of Figure \ref{fig:energy-strain} explicitly show that a prominent instability to a polar $Pna2_1$ phase is present for many perovskite oxides with low-$t$. The same rows in Figure \ref{fig:energy-strain} also demonstrate that the relative phase stability between $Pnma$, $Pna2_1$ and rhombohedral LiNbO$_3$-type $R3c$ (an extremely distorted phase typically formed when $t<0.8$ and characterised by $a^-a^-a^-$ tilts) can be controlled via strain. We do not consider the ilmenite ($R\bar{3}$) structure due to previous reports \cite{parker2011first} indicating that such a structure is higher in energy than the LiNbO$_3$-type structure and this is supported in our calculations using InFeO$_3$ as a representative example - see caption to Table \ref{tab:properties}. Due to the epitaxial constraint forcing the matching of lattice vectors to the substrate, no hexagonal perovskite phases were considered. The $Pmc2_1$ and $Pmn2_1$  symmetries with polarization within the plane were not able to be stabilised under geometry relaxation  \cite{eklund2009strain,yang2012revisiting}.

All materials studied in Figure \ref{fig:energy-strain} show a region of strain in which polar $Pna2_1$ is stabilised over centrosymmetric $Pnma$. For many materials, this region includes 0\%,  indicating that strain is not strictly necessary to stabilise $Pna2_1$. However, there exists an optimum tensile strain at which the energy difference between $Pna2_1$ and $Pnma$ is maximised. Beyond this characteristic optimum strain, the energy difference is reduced until eventually the polarisation is destroyed. This can be seen explicity for ScCrO$_3$, LuMoO$_3$ and YbMoO$_3$ and is expected for the rest of the Sc and In series if higher strains were explored. However, the same behaviour is not observed in the Mg and Zn series - we explain the diversity of responses to strain later in the section.

In addition, we observe that the LiNbO$_3$-type $R3c$ phase (or $Cc$ under bixaxial strain ) is strongly favoured by tensile strain and very rapidly becomes the stable structural polymorph. Nevertheless, the In$B$O$_3$ and $R$MoO$_3$ families all have a window of stability for the $Pna2_1$ phase at relatively low strains. Mg(Mn,Ge)O$_3$ and ZnMnO$_3$ also have a region of stability but exclusively in the compressive regime. Note that we also ran calculations on ZnSnO$_3$ (see Fig \ref{fig:ZnSnO3}), observing similar trends but an $R3c$ ground state at all strains (unlike the results of Ref \cite{kang2017first}).

Two questions are presented by the energy-strain plots in Figure \ref{fig:energy-strain}; what mechanism is responsible for a universal polar distortion in strongly distorted perovskites and what causes the minima in the energy-strain graph? We begin our discussion by investigating the latter. To do this, we explore how the symmetry-adapted modes of the $Pnma$ structure are altered with tensile strain. A representative example is given for InCrO$_3$ in Figure \ref{fig:R4-}a. Whereas the antipolar distortion of the $A$ sites alone ($X_5^-$) and the average of the two tilt modes ($M_2^+$ and $R_5^-$) remains approximately constant, we observe a pronounced minimum in the small antipolar motion of the $A$ and O sites ($R_4^-$) . Of all the modes, tensile epitaxial strain causes the largest relative change in $R_4^-$. In Figure \ref{fig:R4-}b, we show that the $R_4^-$ mode can be decomposed into two components that affect only the $A$ and O-sites respectively. Strain causes continuous but opposite changes in both components, with each changing sign, so that there exists a minimum in the total mode amplitude.  Figure \ref{fig:R4-}b also shows that this mode minimum neatly overlaps with the minimum of the energy of the $Pna2_1$ symmetry.
\begin{figure*}
\centering
\includegraphics[width=\textwidth]{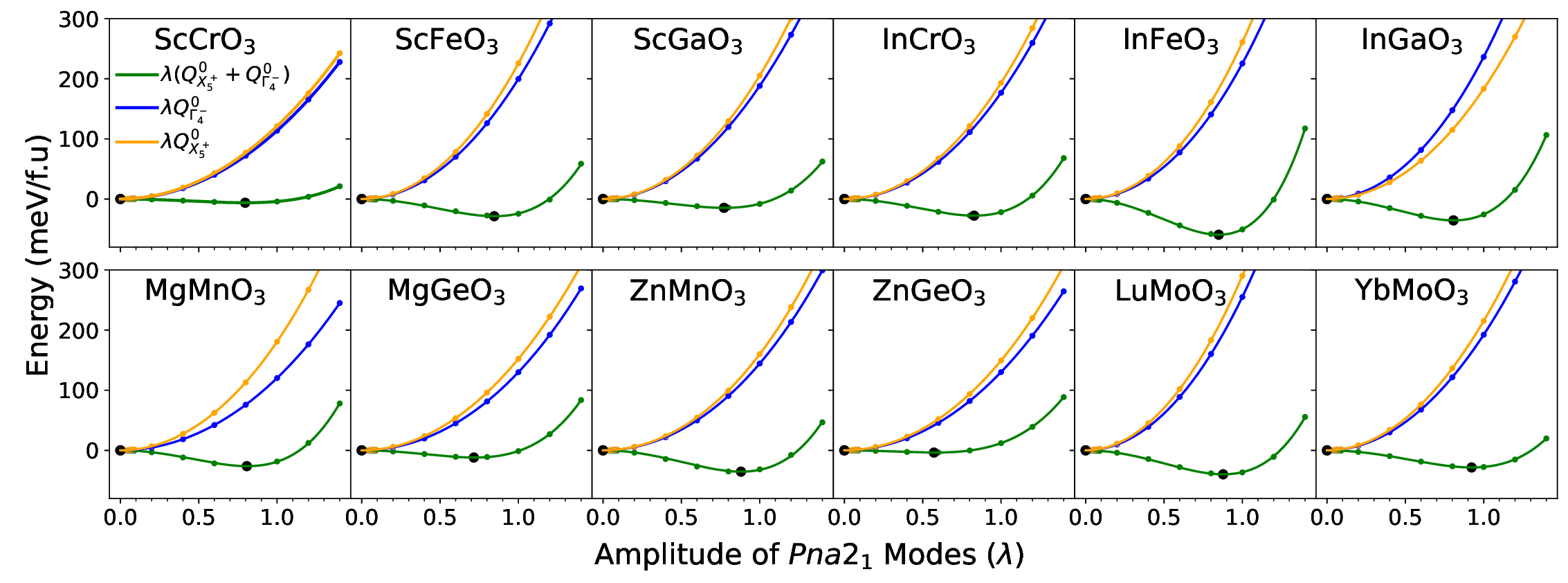} 
\caption{\label{fig:all_avalanche}Energy wells formed by scaling combinations of the modes present in $Pna2_1$ on top of a fully relaxed $Pnma$ structure. Modes are scaled via a parameter $\lambda$. For example, $Q_{\Gamma_4} = {\lambda}Q^0_{\Gamma_4^-}$ where $Q^0_{\Gamma_4^-}$ is the relaxed amplitude of the $\Gamma_4^-$ at 4\% strain (2\% for rare-earth materials). A double well (minima indicated by black points) is formed when both modes are introduced.} 
\end{figure*}

The lowest order coupling terms allowed by symmetry between antiferrodistortive (AFD) modes (such as octahedral tilts and antipolar motions) and polar modes ($\Gamma_4^-$) have the form

\begin{equation}
\label{eq:even_couplings}
\Delta{E}_{\mathrm{even}} = \sum_{i}a_{i,\Gamma_4^-}Q_i^2Q_{\Gamma_4^-}^2,
\end{equation}
where $i$ iterates over the AFD modes present in $Pnma$ and $Q_{\Gamma_4^-}$ is the amplitude of the polar distortion. Benedek and Fennie showed that the coupling constants $a_{i,\Gamma_4^-}$ are large and positive, leading to a suppression of the polar mode \cite{benedek2013there}. In particular, their work highlights the small $R_4^-$ as having a strong, competitive interaction with the polar mode despite its relatively small size. The materials studied in Figure \ref{fig:energy-strain} extend their work to perovskites with even lower $t$, increasing the amplitudes of all AFD modes and subsequently also the importance of the competitive biquadratic interaction. The surprising behaviour of the $R_4^-$ with tensile strain initially leads to a reduction in this interaction, softening the polar mode just enough to lead to a stable $Pna2_1$ phase. With additional increase of strain, the $R_4^-$ mode then increases leading to the hardening of the polar mode and the eventual loss of metastability for the $Pna2_1$ structure. 

Figure \ref{fig:energy-strain} shows identical behaviour in all compounds studied; the minimum of $R_4^-$ coincides with the stability of the $Pna2_1$ phase, supporting the importance of the biquadratic coupling between $R_4^-$ and the polar mode. The exception is for the Zn and Mg series where the energy does not have a minima in the range of strains studied but the $R_4^-$ does. We attribute the lack of an energy minima to the slightly larger tolerance factor in these materials which leads to an overall reduction in all $Pnma$ mode magnitudes and a subsequent decrease in all biquadratic interactions. 

Having explored how the lowest order even terms in the Landau expansion lead to the presence of the $Pna2_1$ energy minima, we note that the inclusion of higher order even terms do not add anything substantial to the analysis. Even-order terms that are quadratic in the polar mode and quartic in the tilts have been calculated to have negative coefficients so that extremely large tilts actually favour polarization in certain perovskites with $R3c$ symmetry \cite{gu2018cooperative}. We show explicity in Figure \ref{fig:minimum tilts} that
this is not the case for $Pna2_1$.

To address the second question and understand the origin of the instability, we  performed further first principles simulations in which we took the fully relaxed $Pnma$ structure at a particular strain ($2\%$ for the rare earth compounds and $4\%$ for the others) and investigated how the energy changes as we introduce structural distortions that break the symmetry to $Pna2_1$. These are the polar mode along the long axis ($\Gamma_4^-$) and the antipolar motion of cations on the $B$ sites ($X_5^+$).  We have not included the other antipolar $B$ site mode ($R_5^+$) introduced in the $Pna2_1$ symmetry because the amplitude in the fully relaxed structures is negligibly small. These results are shown in Figure \ref{fig:all_avalanche}. 

Surprisingly, we see that neither the polar $\Gamma_4^-$ mode nor the antipolar $X_5^+$ are unstable when introduced alone. If the two modes are instead introduced together, a double well forms and both modes obtain a non-zero amplitude. It is apparent that both modes must be present in order to produce a polar structure. This idea was further validated by calculating the dynamical force constants of the $Pnma$ phase in InFeO$_3$ and noticing the existence of an unstable phonon of nearly 50:50 hybrid character. These results also indicate that neither polar $\Gamma_4^-$ or antipolar $X_5^+$ modes are the primary order parameters. Hence, ferroelectricity can not be of proper (unstable $\Gamma_4^-$) type, and indeed we find nominal Born-effective charges as was also observed in $Pna2_1$ fluorides \cite{garcia2016strain}. Nor can ferroelectricity be of the improper (unstable $X_5^+$ driving secondary appearance of $\Gamma_4^-$) type. 

We also rule out the possibility that a negative biquadratic coupling constant between $X_5^+$ and $\Gamma_4^-$ is driving the  simultaneous appearance of $X_5^+$ and $\Gamma_4^-$. In Figure \ref{fig:pos_biquadratic}, we show that this constant is actually positive. We therefore conclude that a triggered mechanism \cite{toledano1979phenomenological} caused by a cooperative biquadratic interaction is not driving the transition to $Pna2_1$.

To explore other possible origins for the $Pna2_1$ instability, we enumerated the lowest odd order couplings in the Landau-like expansion. Terms in which both polar $\Gamma_4^-$ and the antipolar $X_5^+$ are coupled at odd order to modes in the $Pnma$ structure would explain the unusual behaviour of Fig \ref{fig:all_avalanche}, irrespective of the sign of the coefficient. The lowest order terms of this form are

\begin{eqnarray*}
\Delta{E}_{\mathrm{odd}} &&= 
 c_1Q_{X_5^+}Q_{X_5^-}Q_{\Gamma_4^-} + c_2Q_{X_5^+}Q_{R_5^-}Q_{M_2^+}Q_{\Gamma_4^-} \\ +&& c_3Q_{X_5^+}Q_{R_4^-}Q_{M_2^+}Q_{\Gamma_4^-} + c_4Q_{X_5^+}Q_{R_5^-}Q_{M_3^+}Q_{\Gamma_4^-} .
\end{eqnarray*}

Some of these couplings were previously identified in a computational study predicting the $Pna2_1$ symmetry in PbCoO$_3$ \cite{lou2019genetic}. The terms above describe how couplings with the antipolar motion of the A-site ($X_5^-$), the  octahedral tilts ($R_5^-$ and $M_2^+$) and another antipolar mode affecting both the A-site and O-sites ($R_4^-$), all of which have substantial amplitude in small tolerance factor perovskites, can drive the simultaneous appearance of both $\Gamma_4^-$ and $X_5^+$. 

Behaviour of this sort is reminiscent of the avalanche-ferroelectric mechanism studied in Aurivillius compounds\cite{perez2004competing,perez2008multiple,petralanda2017influence}. Here, the condensation of a single mode forces the condensation of two others. In the case of the present small-$t$ perovskites, the large amplitudes of the $Pnma$ modes in the above trilinear and quadlinear couplings enable the condensation of both the polar mode and the antipolar $X_5^+$ mode. Despite the mathematical similarity, whether these low tolerance factor perovskites should be classified as resulting from an \textit{avalanche transition} is not clear. This classification would require that the transition goes directly from the high symmetry cubic $Pm\bar{3}m$ to the low-symmetry $Pna2_1$, While this might occur under certain conditions (e.g. at a particular strain and composition), it seems more likely that the octahedral tilts would condense first to produce $Pnma$ and then at a lower temperature, driven by the above couplings, the $Pna2_1$ phase is stabilised, though this is beyond the scope of this study. This second phase transition could be described as triggered-like since the trilinear (and quadlinear) couplings can renormalize the quadratic terms \cite{perez2004competing,toledano1987landau}.  
\begin{figure*}
\centering
\includegraphics[width=\textwidth]{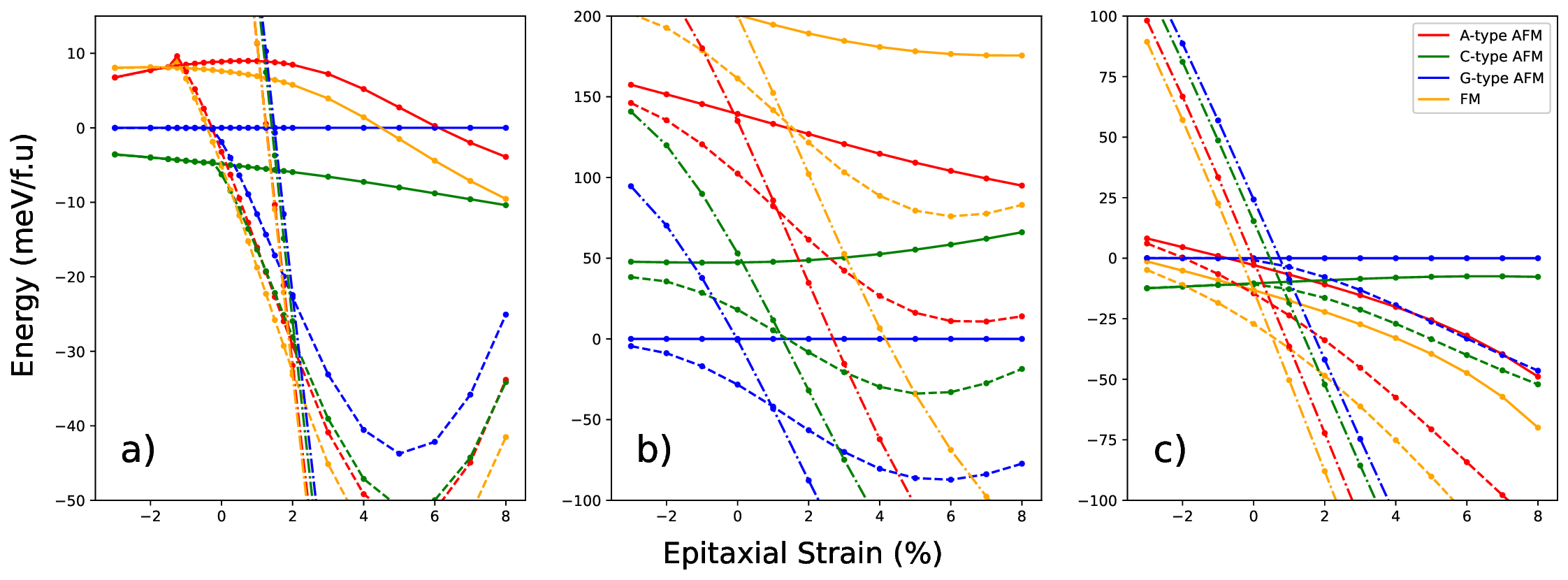}% Here is how to import EPS art
\caption{\label{fig:magnetic}Strain control of magnetic and crystal structure in a) InCrO$_3$, b) InFeO$_3$ and c) MgMnO$_3$. Solid line = $Pnma$. Dashed line = $Pna2_1$. Dash-dot line = $R3c$. Energies with respect to $Pnma$ $G$-type AFM.}
\end{figure*}

To position our work in the context of the existing literature, the $Pna2_1$ phase is relatively rare in perovskite oxides and there is currently much debate into the causes. The phase has been observed in lone-pair systems like BiInO$_3$ \cite{belik2006biino3}, PbRuO$_3$ \cite{cheng2013anomalous} and predicted in PbCoO$_3$ \cite{lou2019genetic}. It has similarly been identified in various $d^0$ materials like CdTiO$_3$ \cite{sun1998study,moriwake2011first,kang2017first}. $Pna2_1$ symmetry has also been observed in rare earth orthoferrites and orthochromates \cite{oliveira2020local,zhao2019first} but is usually ascribed to a spin-driven symmetry breaking, although conflicting reports of the $Pna2_1$ symmetry appearing at a much higher temperature than the rare earth $T_N$ also exist \cite{mishra2023structural}. While lone-pair, $d^0$ and spin-driven effects might be important in certain cases, we argue that the mechanism presented in the present study must also be present and is universal to all low-$t$ systems since it depends solely on the symmetry of the parent phase.

We conclude that $Pnma$ materials are driven to the $Pna2_1$ symmetry by odd order terms such as the trilinear and quadlinear terms discussed above. However, these terms favouring $Pna2_1$ are in contest with the even-order biquadratic terms which inhibit any polarization. Tensile strain manages to reduce the amplitude of the the $R_4^-$ mode so that its corresponding competitive interaction with the polar modes is lessened and the odd-order terms dominate producing the $Pna2_1$ symmetry. This promotion of an out-of-plane polarization in the $Pna2_1$ phase with tensile in-plane strain is opposite to the usual trend found in perovskites which would favour in-plane and disfavour out-of-plane polarization. Further strain starts to increase the magnitude of the $R_4^-$ mode leading to the destruction of the polar phase. The crucial contribution of the $R_4^-$ mode is particularly pronounced in small tolerance factor perovskites because this mode is larger in these materials, reaching $R_4^-$ amplitudes of 0.3 {\AA} or more, considerably larger than the amplitudes of around 0.2 {\AA} or less achieved in the higher $t$ Mg and Zn series or those explored in previous studies \cite{benedek2013there}.

\subsection{\label{candidates}Candidate Multiferroic Systems}

The $Pna2_1$ phase appears to be stable despite several of the materials under investigation being magnetic, allowing for the design of new multiferroics. Figure \ref{fig:magnetic} shows the effect of magnetic ordering on the relative energies of the $Pnma$, $Pna2_1$ and the $R3c$ phases as a function of epitaxial strain for three candidate multiferroics: InCrO$_3$, InFeO$_3$ and MgMnO$_3$. 

For InCrO$_3$, we find the ground state $Pnma$ magnetic order to be C-type, in agreement with experiment \cite{belik2012crystal,ding2017unusual}. Interestingly, the critical strain at which $Pnma$ transitions to $Pna2_1$ changes substantially depending on which magnetic structure is used. For $G$ and $C$-type magnetic structures, which have antiferromagnetic spins within each layer of the perovskite cell, the critical strain is approximately $0\%$. For $A$-type and ferromagnetically aligned spins, the critical strain is instead around $-1\%$. We have seen in Section \ref{generality}, that the antipolar motion at the B site contributes to the quadlinear and trilinear invariants. The difference in critical strain could be due to a spin-phonon effect in which the ferromagnetic alignment of intralayer spins softens the antipolar B site motion. Due to this spin-phonon coupling, it should be possible to also engineer the $Pna2_1$ phase with a compressive strain under an externally applied magnetic field.

Owing to this lower critical strain, ferromagnetism is the magnetic ground state in the region where $Pna2_1$ is stable. Investigating $Pna2_1$ at $1\%$ strain, reveals a band gap of $E_g = 1.84 \ \mathrm{eV}$ - a rare ferromagnetic insulator.
At $1\%$ strain, the polarization is $11.6$ ${\mu}$C/cm$^2$ and the energy difference between the two ferromagnetic orthorhombic structures is ${\Delta}E_O = 29.7\ \mathrm{meV}/\mathrm{f.u}$  whilst ${\Delta}E_R = 317.4\ \mathrm{meV}/\mathrm{f.u}$ between the two rhombohedral ($R\bar3c$ and $R3c$) structures. We use these values as a proxy for switching barrier height and predict that the orthorhombic structures have considerably smaller barriers than the rhombohedral materials, providing a potential method to sidestep the high barriers found in $R3c$ materials \cite{ye2016ferroelectricity,ye2017domain}. 

Figure \ref{fig:magnetic}b shows the same calculation for InFeO$_3$. We do not observe such prominent spin-phonon coupling (possibly due to the presence of $e_g$ orbitals on the Fe$^{3+}$ cation), but the lower barrier present in the orthorhombic structure, presence of high $T_C$ Fe$^{3+}$, weak ferromagnetic canting and the out-of-plane ferroelectricity make InFeO$_3$ a potentially useful thin film multiferroic.  MgMnO$_3$ (Figure \ref{fig:magnetic}c) behaves analogously to InCrO$_3$, likely due to the analogous $d^3$ filling in both,  and also displays a ferromagnetic-insulating state over a substantial range of strains. A summary of all calculated material properties is included in Table \ref{tab:properties} of the supplemental information. The supplemental information also illustrates how our results are altered by the strength of electronic correlation - the existence of a $Pna2_1$ instability is robust to correlation effects but the strains at which it is achieved are altered. 

To conclude, through first principles calculations and group theoretical analysis, we have explored the key role played by the couplings between AFD modes in the $Pnma$ structure to polar distortions in stabilising the technologically useful $Pna2_1$ phase with an out-of-plane polarization. Couplings like these lead to an unusual avalanche-like transition in small tolerance factor perovskites. This mechanism appears resistant to the $d^0$ rule usually prohibiting the creation of novel multiferroic materials. Using this idea, we identify InFeO$_3$ as a potential high $T_C$ multiferroic with a small polarization switching barrier  and sizeable wFM moment. We further identify InCrO$_3$ and MgMnO$_3$ as polar, ferromagnetic insulators exhibiting strong spin-phonon coupling.

\section{Methods}
All simulations are performed using density functional theory (DFT) as implemented in the Vienna Ab-Initio Software Package (VASP) Version 6.3.2 \cite{blochl1994projector,kresse1996efficiency,kresse1996efficient,kresse1999ultrasoft}. We use the Perdew-Burke-Ernzerhof exchange correlation functional for solids (PBESol)\cite{perdew2008restoring}. We use a high plane wave cutoff energy of 800 eV to ensure convergence for all systems studied as well as a 7x5x7 Monkhorst-Pack $k$-grid for the $\sqrt{2}{\times}2{\times}\sqrt{2}$ 20-atom supercell. Self consistent field calculations were continued until differences in energies were within a tolerance of $10^{-8}$ eV. Geometry relaxations were continued until the smallest Hellman-Feynman force was less than $10^{-3}$ eV/\AA. We use projector augmented wave pseudopotentials in all our calculations. A summary of which electrons are treated as valence is presented in Table \ref{tab:electrons}. To better approximate the effect of electron localization and correlation, we use the rotationally invariant formulation of the onsite Hubbard-$U$ parameter \cite{dudarev1998electron}. We use a consistent value of $U=4$ eV for all materials with unpaired $d$ electrons and $U=0$ eV for those without. This is also the methodology employed in previous computational screening studies \cite{zhao2019first}. The results are not qualitatively affected by the choice of $U$ - see supplemental information for details.

To apply epitaxial strain to our systems, we utilize an additional patch to VASP \cite{ioptcell} that fixes selected lattice constants during a relaxation. This simulates epitaxial growth on a substrate for a perovskite film with many layers.

The symmetry analysis was conducted using the INVARIANTS tool of the ISOTROPY Software Suite \cite{isodistort,campbell2006isodisplace}.

\section{Data Availability}
All data supporting the plots within this paper and other findings of this study are available from the Supplementary Information or the corresponding authors upon reasonable request.

\bibliographystyle{apsrev4-2}
\bibliography{biblio}

\section{Acknowledgements}
CAMS and NCB would like to thank Helen He for helpful discussions relating to this research. This work used the Hamilton HPC service at Durham University. CAMS and NCB acknowledge the Leverhulme Trust for a research project grant (Grant No. RPG-2020-206).

\section{Author Contributions}
CAMS performed all computational work. NCB supervised the project. Both authors contributed to the writing of the manuscript.

\renewcommand{\thefigure}{S\arabic{figure}}
\renewcommand{\thetable}{S\arabic{table}}
\setcounter{figure}{0} 
\setcounter{table}{0} 
\clearpage
\onecolumngrid
\begin{center}
  \centering
  \LARGE Universal polar instability in highly orthorhombic perovskites - Supplemental Information \par
\end{center}

\begin{center}
  \centering
  \large Cameron A.M. Scott  \quad Nicholas C. Bristowe \par
\end{center}

\begin{center}
  \centering
  \large Centre for Materials Physics, Durham University, South Road, Durham DH1 3LE, United Kingdom \par
\end{center}

\begin{table*}[b]
\centering
\begin{tabularx}{0.8\textwidth}{c|c|c|c|c|c}
\textbf{Material} & $t$    & \textbf{Magnetism} & \textbf{$a_{IP}$ (\AA)} & \textbf{Relaxed $Pna2_1$ $X_5^+$ (\AA)} & \textbf{Relaxed $Pna2_1$ $\Gamma_4^-$ (\AA)} \\ \hline
ScCrO$_3$         & 0.7539 & C-type AFM                                   & 3.678                 & 0.549                             & 0.854                                  \\ \hline
ScFeO$_3$         & 0.7426 & G-type AFM                                   & 3.670                 & 0.814                             & 1.172                                  \\ \hline
ScGaO$_3$         & 0.7520 & NM                                           & 3.659                 & 0.706                             & 1.029                                  \\ \hline
InCrO$_3$         & 0.7737 & C-type AFM                                   & 3.741                 & 0.728                             & 0.990                                  \\ \hline
InFeO$_3$         & 0.7620 & G-type AFM                                   & 3.734                 & 0.899                             & 1.152                                  \\ \hline
InGaO$_3$         & 0.7717 & NM                                           & 3.715                 & 0.847                             & 1.121                                  \\ \hline
MgMnO$_3$         & 0.7786 & FM                                           & 3.574                 & 0.715                             & 1.164                                  \\ \hline
MgGeO$_3$         & 0.7786 & NM                                           & 3.573                 & 0.684                             & 1.147                                  \\ \hline
ZnMnO$_3$         & 0.7861 & FM                                           & 3.577                 & 0.815                             & 1.055                                  \\ \hline
ZnGeO$_3$         & 0.7861 & NM                                           & 3.568                 & 0.732                             & 0.997                                  \\ \hline
LuMoO$_3$        & 0.7664 & G-type AFM                                   & 3.907                 & 0.766                             & 1.087                                  \\ \hline
YbMoO$_3$        & 0.7688 & G-type AFM                                   & 3.918                 & 0.657                             & 0.947                                 
\end{tabularx}
\caption{Parameters of materials studied. Tolerance factor $t$ measured using 6-coordinated A-sites, giving consistency with (Belik, 2014). In plane lattice constant $a_{IP}$ taken as the average bulk $Pnma$ in-plane lattice constants divided by $\sqrt{2}$. Amplitudes of $Pna2_1$ mode, with respect to the $Pnma$ supercell, calculated at 4\% strain for non rare-earth materials. For rare-earth materials, the calculation is performed at 2\% strain - approximately the minimum of the $Pna2_1$ well. Magnetic structure determined by calculating the energies after bulk relaxation of A-, C-, G- and FM and selecting the lowest energy. }
\label{tab:parameters}
\end{table*}

% \begin{figure}[b]
% \centering
% \includegraphics[width=0.8\textwidth]{new_tilts.eps}% Here is how to import EPS art
% \caption{\label{fig:tilts} Demonstration of the linear relationship between the $X_5^-$ mode and the product of the two tilt modes $R_5^-M_2^+$. This linear relationship is expected from the existence of the trilinear coupling $X_5^-{\oplus}R_5^-{\oplus}M_2^+$.}
% \label{fig:sup-linear}
% \end{figure}

\begin{figure*}[b]
\centering
\includegraphics[width=0.6\textwidth]{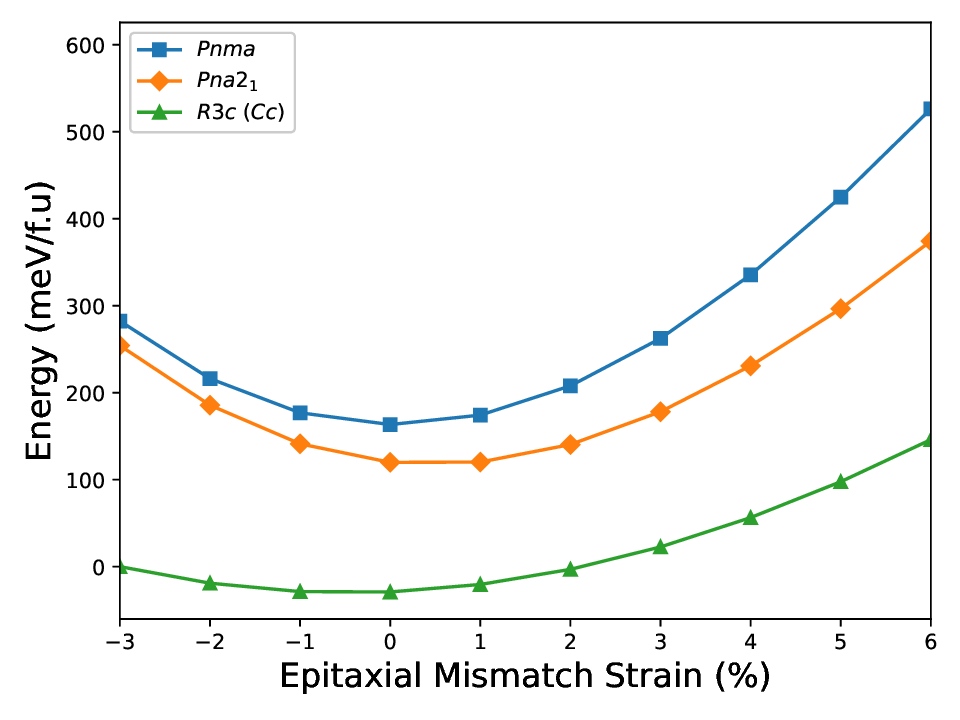}% Here is how to import EPS art
\caption{\label{fig:ZnSnO3}Relative stability, as a function of strain, of $Pnma$, $Pna2_1$ and $R3c$ phases in ZnSnO$_3$. We observe that $R3c$ is the ground state for all values of strain, in contrast to the results of (Kang, 2017).}
\end{figure*}

\begin{figure*}
\centering
\includegraphics[width=0.6\textwidth]{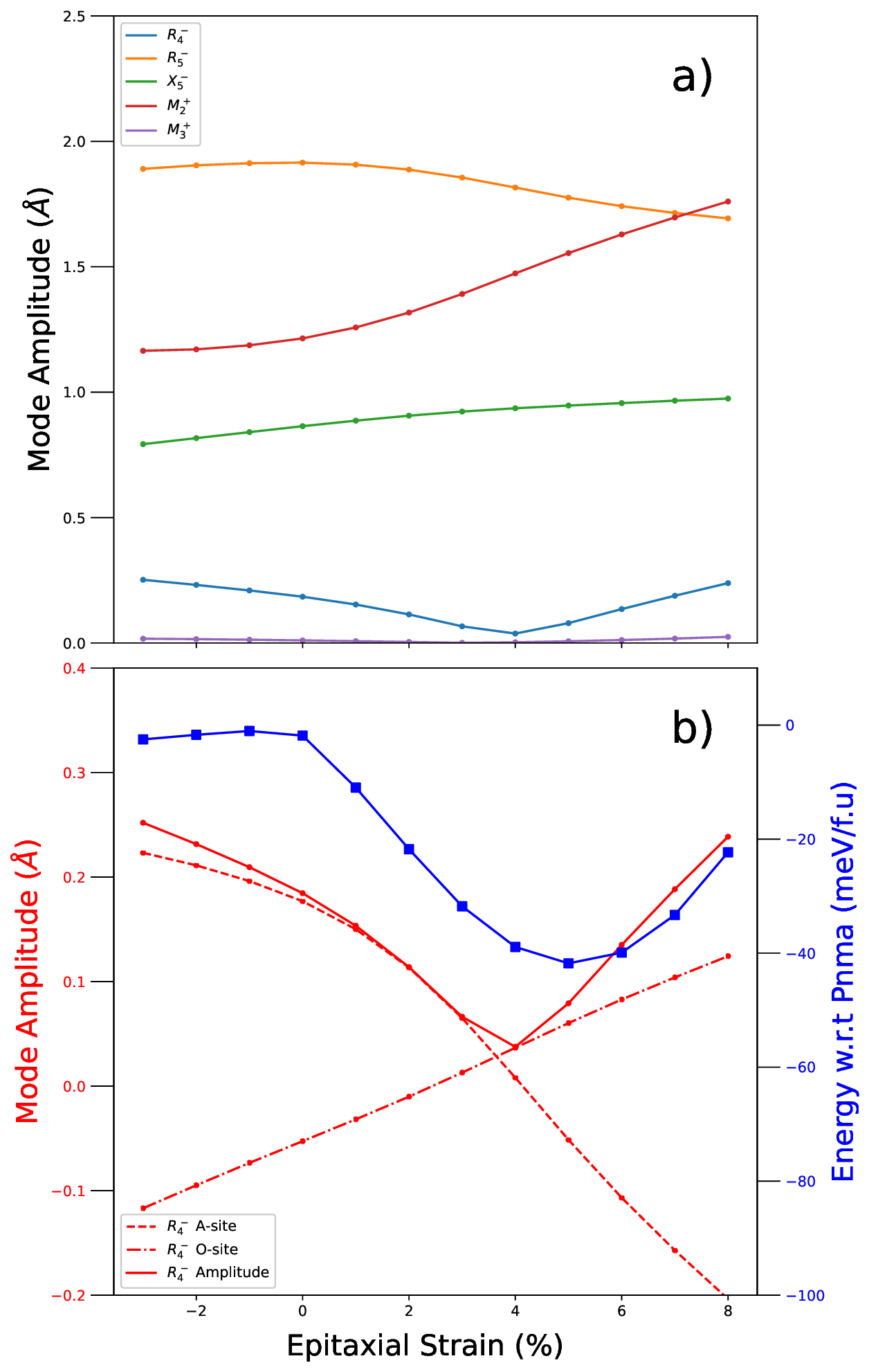}% Here is how to import EPS art
\caption{\label{fig:R4-}a) Amplitude of modes in $Pnma$ InCrO$_3$. We see a pronounced minimum in the amplitude of the $R_4^-$ mode b) Minimum in $R_4^-$ is caused by opposite trends in its two components. Minimum of $R_4^-$ also coincides closely with the minimum of the $Pna2_1$ energy.}.
\end{figure*}

\begin{figure*}[b]
\centering
\includegraphics[width=0.8\textwidth]{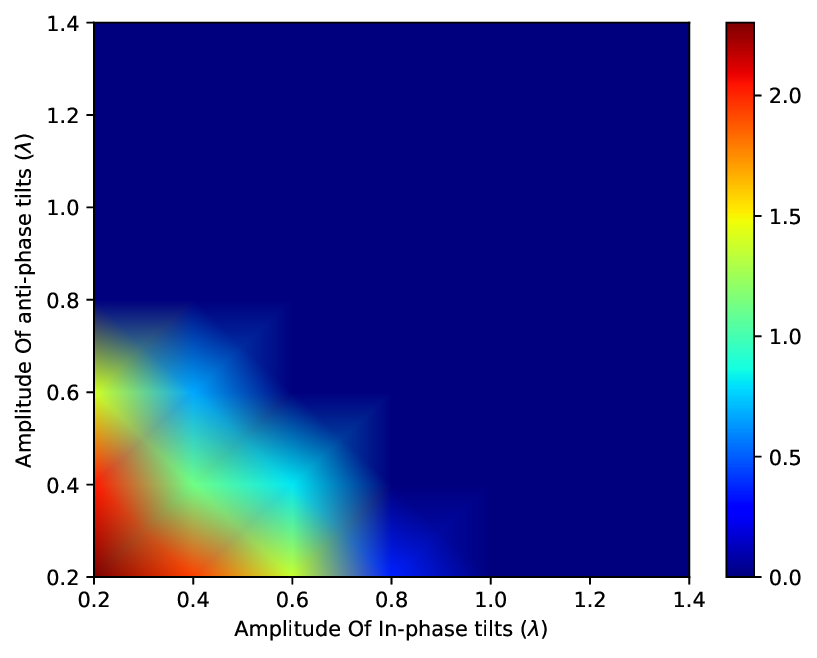}% Here is how to import EPS art
\caption{Plot of the amplitude of the polar distortion of 1\% epitaxially strained InFeO$_3$. Colour scale denotes the amplitude of the polar distortion, defined as the value of $Q_{\Gamma_{4^-}}$ at which the energy is minimised. The deep blue colour for large tilts indicates that the polar mode is not stable when the tilts become large.}
\label{fig:minimum tilts}
\end{figure*}

\begin{figure*}[b]
\centering
\includegraphics[width=0.8\textwidth]{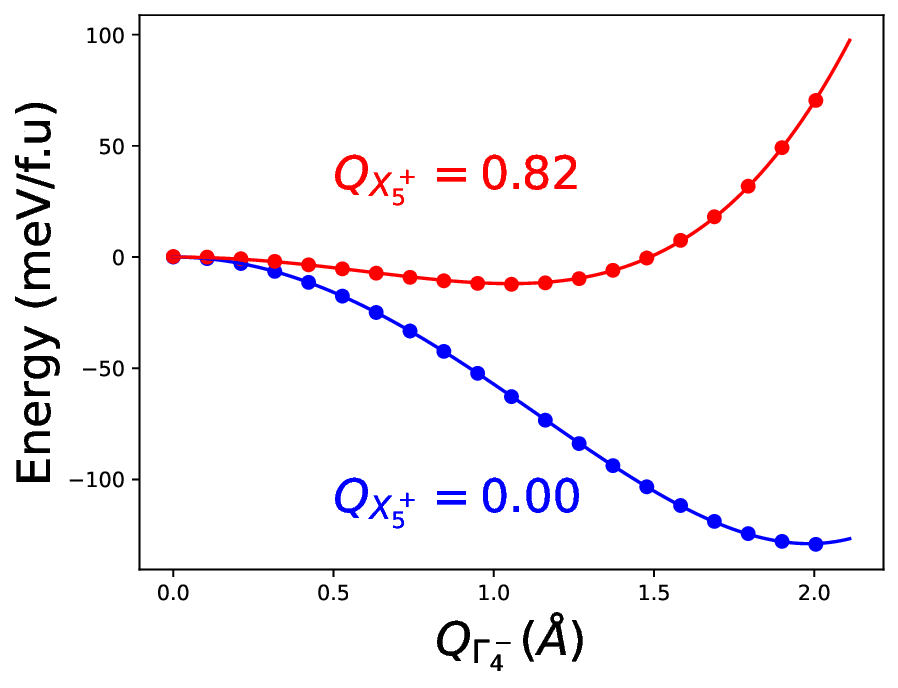}% Here is how to import EPS art
\caption{\label{fig:pos_biquadratic} By introducing the polar $\Gamma_4^-$ mode to the cubic $Pm\bar{3}m$ perovskite cell whilst simultaneously including a fixed magnitude of an antipolar $X_5^+$ distortion, we demonstrate the positive biquadratic coupling between $Q_{\Gamma_4^-}$ and $Q_{X_5^+}$. A positive biquadratic coupling eliminates the possibility of a triggered-like ferroelectric mechanism between these two modes.}
\end{figure*}

\begin{figure*}[b]
\centering
\includegraphics[width=0.8\textwidth]{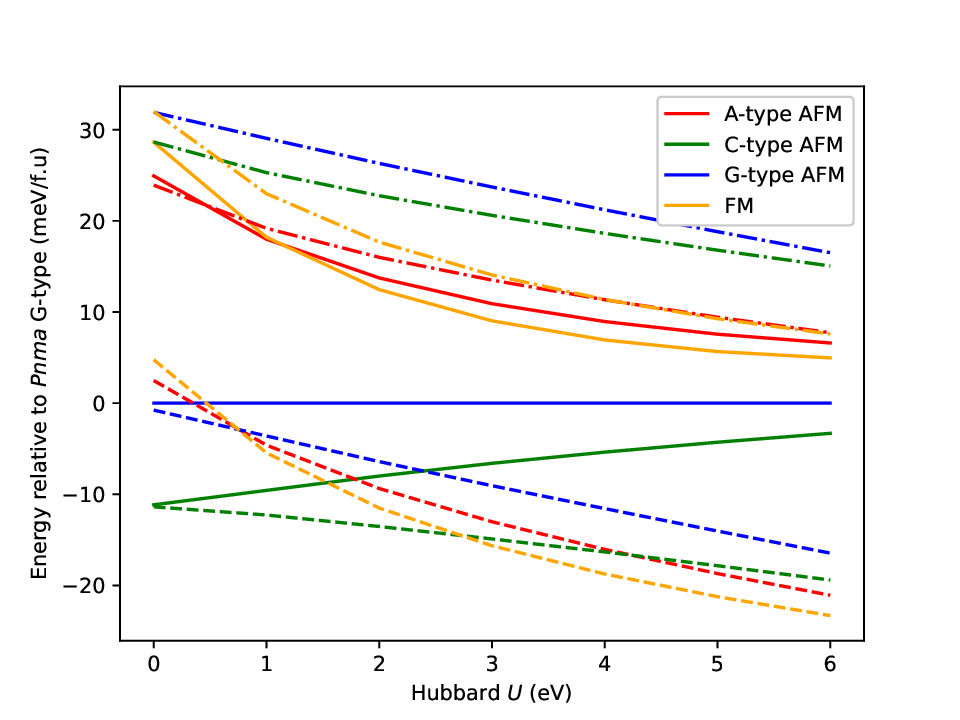}% Here is how to import EPS art
\caption{\label{fig:sup-InCrO3-U} Crystal and magnetic structure of InCrO$_3$ as function of $U$ at 1\% strain. Bold lines denote magnetic structure with $Pnma$ crystal symmetry, dashed lines with $Pna2_1$ crystal symmetry and dashdot lines with $R3c$ crystal symmetry. Lowest line denotes lowest energy crystal and magnetic structure for that particular choice of Hubbard-$U$ }
\end{figure*}

\begin{figure*}[b]
\centering
\includegraphics[width=0.8\textwidth]{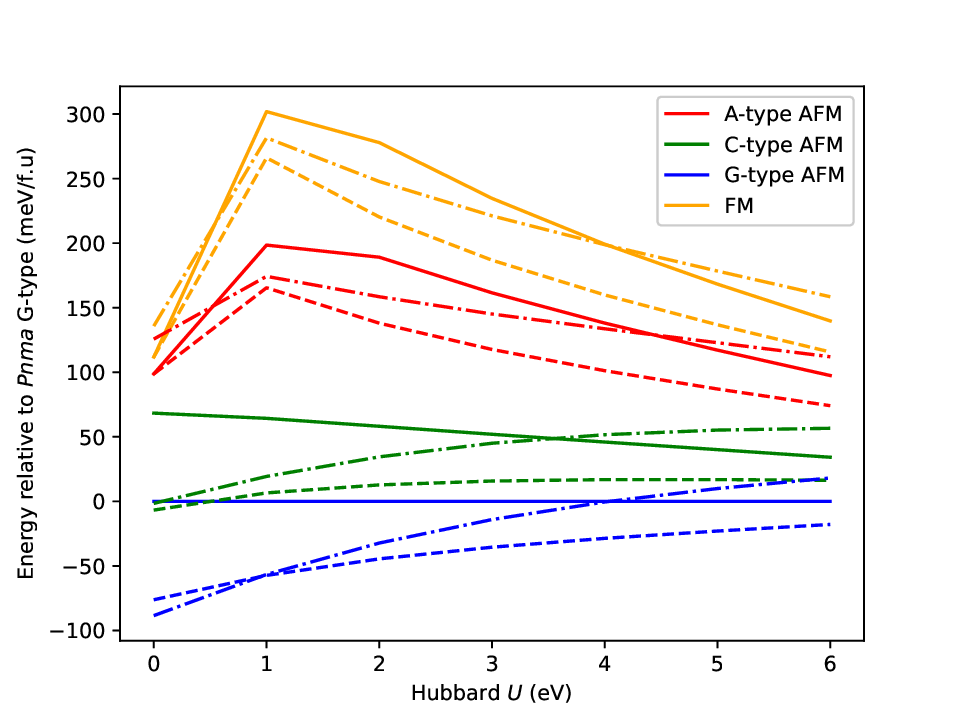}% Here is how to import EPS art
\caption{\label{fig:sup-InFeO3-U} Crystal and magnetic structure of InFeO$_3$ as function of $U$ at 0\% strain.Bold lines denote magnetic structure with $Pnma$ crystal symmetry, dashed lines with $Pna2_1$ crystal symmetry and dashdot lines with $R3c$ crystal symmetry. Lowest line denotes lowest energy crystal and magnetic structure for that particular choice of Hubbard-$U$}
\end{figure*}

\begin{figure*}[b]
\centering
\includegraphics[width=0.8\textwidth]{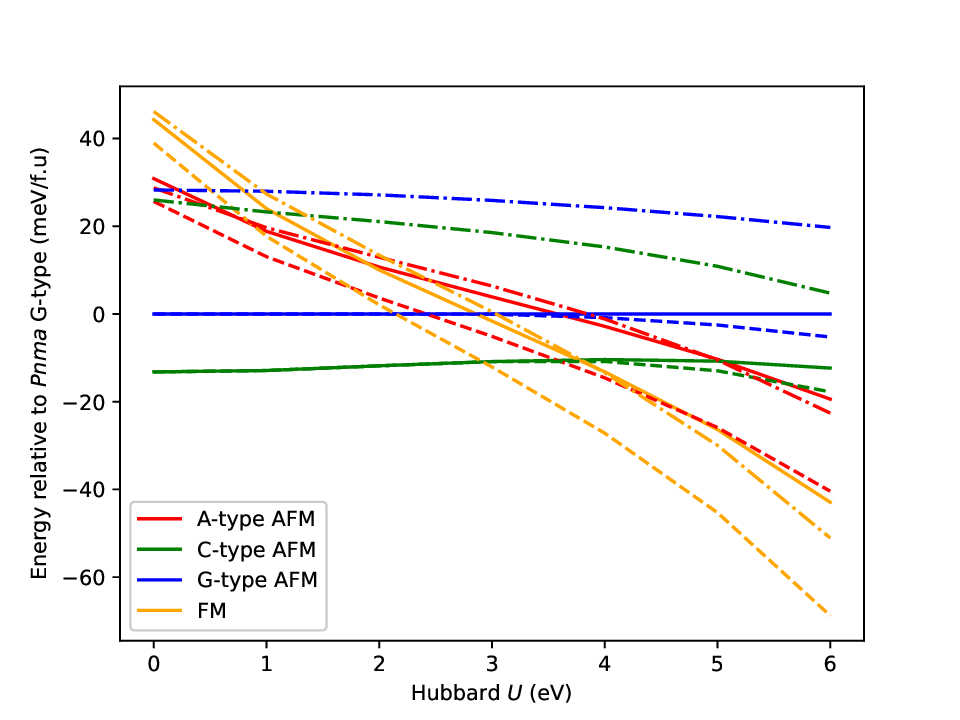}% Here is how to import EPS art
\caption{\label{fig:sup-MgMnO3} Crystal and magnetic structure of MgMnO$_3$ as function of $U$ at 0\% strain.Bold lines denote magnetic structure with $Pnma$ crystal symmetry, dashed lines with $Pna2_1$ crystal symmetry and dashdot lines with $R3c$ crystal symmetry. Lowest line denotes lowest energy crystal and magnetic structure for that particular choice of Hubbard-$U$}
\end{figure*}

\begin{table*}[]
\centering
\begin{tabular}{c|c|c|c|c|c|c|c}
Material  & \begin{tabular}[c]{@{}c@{}}Strain \\  (\%)\end{tabular} & \begin{tabular}[c]{@{}c@{}}Polarization \\ (${\mu}\mathrm{C}/\mathrm{cm}^2$)\end{tabular} & \begin{tabular}[c]{@{}c@{}}Magnetic \\ Structure\end{tabular} & \begin{tabular}[c]{@{}c@{}}Easy \\ Axis\end{tabular} & \begin{tabular}[c]{@{}c@{}}Band \\ Gap \\ (eV)\end{tabular} & \begin{tabular}[c]{@{}c@{}}$\Delta{E}_0$ \\ (meV/f.u)\end{tabular} & \begin{tabular}[c]{@{}c@{}}$\Delta{E}_R$\\  (meV/f.u)\end{tabular} \\ \hline
InCrO$_3$ & 1                                                       & 11.15                                                                   & FM                                                            & [100]                                                & 1.84                                                        & 29.70                                                            & 317.40                                                           \\
InFeO$_3$ & 0                                                       & 22.42                                                                   & G-AFM*                                                        & [001]                                                & 1.81                                                        & 28.30                                                            & 252.96                                                           \\
MgMnO$_3$ & 0                                                       & 15.49                                                                   & FM                                                            & [001]                                                & 0.85                                                        & 13.95                                                            & 331.01                                                          
\end{tabular}
\caption{Properties of candidate materials at specified strains. InFeO$_3$ is a G-type antiferromagnetic - the asterisk denotes a magnetic point group allowing for a wFM moment. This has a calculated magnitude of $0.021\mu_B$. ${\Delta}E_O$ and ${\Delta}E_R$ are the difference in energies between the two orthorhombic structures ($Pnma$ and $Pna2_1$) and the two rhombohedral structures ($R\bar{3}c$ and $R3c$) respectively. We use these values as a proxy for ferroelectric switching barrier height. Note that the ilmenite phase is not considered in the current study. Calculations on this phase in InFeO$_3$ reveal the ilmenite structure to be $502$ meV/f.u higher in energy than the $Pnma$ phase.}
\label{tab:properties}
\end{table*}

\begin{table*}[]
\centering
\begin{tabular}{c|c}
\textbf{Element} & \textbf{Valence Electrons}         \\ \hline
Sc               & $3s^23p^64s^23d^1$                     \\ \hline
Cr               & $3p^64s^13d^5$                     \\ \hline
Fe               & $3d^64p^2$                         \\ \hline
Ga               & $3d^{10}4s^24p^1$                    \\ \hline
In               & $4d^{10}5s^25p^1$                    \\ \hline
Mg               & $3s^2$                             \\ \hline
Mn               & $3p^63d^54s^2$                     \\ \hline
Ge               & $3d^{10}4s^24p^2$                    \\ \hline
Zn               & $3d^{10}4s^2$                        \\ \hline
Lu*              & $5p^65d^16s^2$                     \\ \hline
Mo               & $4s^24p^64d^55s^1$                 \\ \hline
Yb*              & $5p^64f^16s^2$                     \\ \hline
O                & $2s^22p^4$                        
\end{tabular}
\caption{Valence electrons used in our projector augmented wave ultrasoft-pseudopotential calculations. Asterisks mark rare earth elements for which most or all of the highly localized $f$ electrons have been confined to the core}
\label{tab:electrons}
\end{table*}

\end{document}